Nguyen Luong[1], MSc; Gloria Mark[2], PhD; Juhi Kulshrestha[1], PhD; Talayeh Aledavood[1], PhD

[1]Computer Science Department, Aalto University, Espoo, Finland
[2]Informatics Department, University of California, Irvine, Irvine, CA, United States

**Corresponding Author**:
Nguyen Luong, MSc
Computer Science
Aalto University
02150 Espoo
Finland
Phone: +358 442404485
Email: nguyen.luong@aalto.fi

# Sleep during COVID-19 pandemic: Longitudinal observational study combining multisensor data with questionnaires

## Abstract

**Background:**
The COVID-19 pandemic led to various containment strategies, such as work-from-home policies and reduced social contact, which significantly altered people's sleep routines. While previous studies have highlighted the negative impacts of these restrictions on sleep, they often overlook a comprehensive perspective that incorporates other factors that can also influence sleep, such as seasonal variations and physical activity.

**Objective:**
Our study aims to longitudinally examine the fine-grained changes in sleep patterns of working adults during the COVID-19 pandemic using a combination of repeated questionnaires and high-resolution passive measurements from wearable sensors. We investigate the association between sleep and 5 sets of variables capturing characteristics of individuals including (i) demographics, (ii) sleep-related habits, (iii) physical activity behaviors, and external factors including (iv) pandemic-specific constraints, and (v) seasonal variations during the study period.

**Methods:**

We recruited working adults in Finland to participate in our one-year-long study (June 2021 - June 2022) conducted in the late stage of the COVID-19 pandemic. We collected multisensor data from fitness trackers worn by the participants as well as work and sleep-related measures through monthly questionnaires. Additionally, we used the stringency index for Finland at different points in time to estimate the degree of the pandemic-related lockdown restrictions in place during the study duration. We applied linear mixed models to study the changes in sleep patterns during the late stage of the pandemic and their association with the aforementioned five sets of variables.

**Results:**

The sleep structure of 27,350 nights from 112 working adults was analyzed. We observed that more stringent pandemic measures were associated with longer total sleep time (TST) ($\beta$= 0.02, 95% CI 0.01 to 0.03, P<.001) and later midpoint of sleep (MS) ($\beta$= 0.1, 95% CI 0.09 to 0.11, P<.001). Being a snoozer (i.e. snoozing the alarm when waking up from sleep) was associated with higher variability in both TST ($\beta$= 0.34, 95% CI 0.13 to 0.54, P=.002) and MS ($\beta$= 0.24, 95% CI 0.07 to 0.41, P=.006). Service staff slept more than academic staff ($\beta$= 0.3, 95% CI 0.09 to 0.51, P=.005) with also lower variability in TST ($\beta$= -0.29, 95% CI -0.49 to -0.08, P=.006). Engaging in physical activity earlier in the day correlated with longer sleep duration ($\beta$= 0.06, 95% CI 0.05 to 0.08, P<.001).

**Conclusions:**

Our study provided a comprehensive view of the possible factors affecting sleep patterns during the late stage of the pandemic. Our results revealed that stringent measures implemented during the COVID-19 pandemic are associated with changes in sleep patterns. The more flexible work-life routine arising from the restriction is also linked with changes in sleep-related habits among different occupations.

**Keywords:**

## Introduction

Sleep is a crucial component of our daily lives, tightly interconnected with all aspects of daily routines and our overall well-being, such as mental health [44, 45], physical health [58], and work performance [57, 55]. The COVID-19 pandemic had a deep impact on various aspects of people's daily lives, with one particularly significant area being sleep patterns. However, the pandemic's effects on sleep were often indirect, arising from changes in daily routines and lifestyle adjustments, rather than as a direct consequence of the virus.

As a response to the pandemic, outdoor restrictions limited our exposure to natural daylight, a crucial element known to regulate our circadian rhythms and sleep patterns [39]. Similarly, mobility restrictions altered the structure of daily physical activity (PA). Additionally, workplace restrictions resulted in work-from-home policies, which led to reduced mobility and flexible working hours. While all of this led to more relaxed work schedules, it also blurred the boundaries between professional and personal life. Notably, all these factors - daylight exposure, physical activity, and work routine - all significantly affected by the pandemic, are well-established influences on sleep health [53, 65]. Therefore, a comprehensive view of sleep, accounting for all these variables in the context of the pandemic, is essential to fully comprehend the extent and nature of its impact.

Sleep measurements traditionally rely on self-reported methods, such as the Karolinska [36] or the Pittsburgh sleep diary [35]. While those methods are suitable for tracking day-to-day sleep over a few days or several weeks, conducting diary studies over longer time intervals is usually not favorable due to the cognitive burden on the participants. Non-intrusive measurements using smartphones and fitness trackers have recently emerged as a more viable alternative for capturing sleep over extended periods. While the consumer-grade devices remain incapable of precisely detecting sleep stages, for sleep onset, duration, and wake-up time, these devices have shown more promising results. Assessment of sleep through these devices has the advantage of measuring sleep in people's natural living environments as opposed to sleep laboratories and is not affected by memory biases, which can occur with survey responses and sleep diaries. Prior studies have shown that wearable devices have proven to be valuable tools for measuring sleep patterns in various use cases. For instance, they have been utilized to determine people's chronotypes and track their sleep and activity rhythms over extended periods [43, 47]. These devices have also been employed to measure sleep alignment between coworkers [41], examine the relationship between sleep and burnout [46], and assess sleep patterns in different populations, including patients with mental disorders [42]. Several studies have confirmed the validity and reliability of those wearables, demonstrating respectable sensitivity against the gold-standard polysomnography (PSG). For instance, a review of seven consumer sleep-tracking devices [37] has demonstrated their high effectiveness in detecting sleep with respect to PSG. Similarly, a study [68] evaluated six consumer wearable devices and confirmed their validity in assessing the timing and duration of sleep when compared to PSG.

Prior research has compared sleep patterns before and during the pandemic, revealing notable differences. Studies found that after the onset of the pandemic individuals tended to go to bed later [1], sleep for extended durations [2], exhibited smaller variations between weekday and weekend sleep [32, 27], and experienced increased sleep disturbances or diminished sleep quality [4]. Various factors are identified as contributing to these sleep routine disruptions, including a decrease in physical activity [3], social isolation [69], increased usage of electronic devices [4], and the ability to work from home [41].

While previous studies have focused on the immediate consequences of the lockdowns and restriction policies, less attention has been given to the long-term effects, especially in the late stages of the pandemic when restrictions began to relax. This is an important phase to study as it can provide knowledge on the residual effects of the pandemic on sleep patterns and how quickly people revert to their pre-pandemic sleep habits. The transition to work-from-home as the default working mode has resulted in a less constrained work-life routine, which leads to the tendency to maintain a flexible sleep-wake schedule. Certain demographics could better leverage these transitions, such as individuals with more flexible routines like research personnel, or people who tend to snooze their alarms after waking up from sleep, which we refer to as "snoozers".

Additionally, occupation is a known factor influencing sleep patterns, with the classic example being the contrast between shift workers and non-shift workers [24, 56]. However, there is less understanding regarding the disparities between various roles within academia, such as researchers with deadline-driven roles and administrative personnel typically following a 9-5 schedule. Therefore, a more comprehensive, longitudinal analysis of sleep patterns that includes these variables and extends into the late stages of the pandemic is important.

## Objectives

Our study aims to provide a holistic view of how the pandemic has influenced sleep patterns. In particular, we evaluate the long-term relationships between sleep patterns, including average and variability in total sleep duration and sleep timing, in conjunction with individuals' characteristics (demographics, occupation, physical activity) as well as external factors (stringency of restriction policies, seasons). Our research utilizes longitudinal data gathered from fitness trackers and questionnaires of working adults from a Finnish university. This extensive dataset allows us to examine the shifts in sleep behavior during the later stages of the COVID-19 pandemic, from June 2021 to June 2022.

## Methods

We used the data from our cor:ona (comparison of rhythms: old vs. new) Study [38] to a one-year multimodal data set of working adults. The study was approved by the Aalto University Research Ethics Committee.

## Participants and Procedures

The cor:ona Study recruited 128 full-time employees from a university in Finland for a one-year study to examine how their daily activities changed during different stages of the COVID-19 pandemic. Throughout the study, participants wore a fitness tracker (Polar Ignite) which allowed us to unobtrusively collect various measures related to sleep and physical activity. In addition, they completed an initial baseline and an exit questionnaire, as well as a shorter version of the baseline questionnaire each month. The monthly questionnaires asked participants to provide information about their daily routines, work, and sleep quality during the past month.

**Fitness Tracker Data**

*Sleep measures*

The fitness trackers measure sleep start time (the registered time pointed when a person fell asleep), sleep end time (the registered time pointed when a person woke up), and interruption duration (total time in seconds a person spent awake between sleep start time and end time) for each day [62]. Here, a sleep period was defined as the most extended sleep episode for each day. Sleep patterns were measured using four metrics: (1) Total Sleep Time (TST) measured the time a person spends in bed, determined by the duration from bedtime to waketime, minus the interruption duration (2) Midsleep (MS) point was used to measure sleep timing, computed as the midpoint between bedtime and waketime. In addition, we propose two other metrics as a proxy for sleep regularity: (3) TST variability was computed using the weekday (Sunday night to Thursday night) standard deviation of TST, and similarly (4) MS variability was computed using the standard deviation of MS during weekdays. We only considered weekdays due to the expected differences between weekday and weekend sleep patterns. The Niimpy behavioral data analysis toolbox was used for sleep measurement extraction [64].

*Physical activity measures*

We used step counts captured at hourly intervals from the fitness trackers to measure physical activity. In studies assessing physical activity, it is commonly reduced to a single metric: the overall activity volume. However, to comprehensively account for daily physical activities, including their timing and distribution, we introduced two additional metrics: midstep and steps entropy. These metrics aim to capture the temporal occurrence and dispersion of physical activities throughout the day. Specifically, midstep represents the hour of the day when half the total number of steps is achieved, analogous to midsleep for physical activity. On the other hand, step entropy quantifies the concentration of steps over the day and is measured using Shannon entropy:

$$H(X) = - \sum_{i=0}^{23} P(x_i) \, log_2(P(x_i))$$

in which $P(x_i)$ is the distribution of steps within bin $i$ with $i$ ranging from 0 to 23 (corresponding to the 24 hours of the day). The Shannon entropy measure of physical activity reflects the degree of randomness and uncertainty in activity levels throughout the day. Intuitively, lower entropy values indicate a higher level of predictability, with physical activity concentrated in specific periods. Conversely, higher entropy values indicate a lower level of predictability, with physical activity spread more across the day.

### External Data

Seasonal data was collected from the World Weather Online developer API [31]. Since day length varied significantly in Finland during the course of the study (up to 13 hours), it was used as a proxy for seasonal variables. The choice of day length as a proxy is motivated by [33]. The study contrasted two geographically distinct locations with substantial differences in day length variability (Ghana and Norway). While there were no noticeable seasonal effects of day lengths on Ghanaians, Norwegians demonstrated a delay in bedtime and wake time during summer weekdays, although the sleep duration remained relatively unaffected.

We also utilized the stringency index [8], a composite index ranging from 0 to 100 to measure COVID-19 restriction policies. Higher values on this index reflect the implementation of more rigorous COVID-19 restriction policies, encompassing measures like school and workplace closures, the cancellation of public events, and the enforcement of stay-at-home orders. This index enables standardized comparisons of policy responses across different countries or regions or changes within the same region over time. The index was recorded on a daily basis.

### Questionnaire Data

Upon entering the study, participants were asked to provide basic background information in a baseline questionnaire, including age, gender, chronotype, occupation, and origin, among others. Chronotype was measured using the reduced Morningness-Eveningness Questionnaire (MEQ) [30], with a higher score indicating morning type and a lower score indicating evening type. For the origin-related question, participants were given three choices: Finland, Europe (except Finland), or outside of Europe. We refer to those who indicated they are from Finland as Finnish and others as "migrant background". For questions concerning occupation, the participants were asked whether they were academic or service staff. The term "academic staff" represents individuals involved in academic and research activities within the organization. On the other hand, "service staff" refers to individuals in roles such as human resources, and other administrative or support functions. Snoozing behavior was assessed using the following question: "Snoozing can be considered as choosing to go back to sleep after an alarm has awakened you intending to wake up later; setting an alarm earlier than when you intend to wake up; or setting multiple alarms with the intent to not wake up on the first alarm. Do you currently consider yourself a snoozer using this definition?", as adapted from [14].

For snoozer characteristics analysis, we employed the Patient Health Questionnaire-2 (PHQ-2) [59] and the short form of the Pittsburgh Sleep Quality Index (PSQI) [60], averaging the values collected from the monthly questionnaires. Furthermore, the short form Positive and Negative Affect Schedule (PANAS-SF) [61] was gathered at the initial baseline questionnaire.

### Data exclusion and Preprocessing

Sleep data were restricted to the time range from July 1, 2021, to May 31, 2022, due to our rolling recruitment process that began in mid-June 2021 and ended in June 2022, in order to exclude months where data was not collected for the whole month. A standard filter, adopted from [7], was applied to remove outliers' TSTs (TST < 3 and TST > 13). Participants with less than 30 nights recorded due to dropout or technical issues were excluded. For gender-related analysis, non-binary participants (N=1) were excluded to preserve their privacy.

### Statistical Analysis

Linear mixed models [34] were used to test whether sleep patterns and their regularity changed over time while adjusting for demographics, physical activities, season, and policy stringency. Models included TST, MS, and variability of TST and MS as dependent variables. We adopt a sequential modeling strategy in which we sequentially build three distinct models for each dependent variable. Model 1 consists of basic characteristics such as chronotype, age, gender, origin, occupation, and cohabitation status. Model 2 extends Model 1 by adjusting for external factors such as the stringency of the restrictions and day length. Finally, Model 3 extends Model 2 by adding physical activity metrics using the number of step count, midstep, and step entropy. This approach allows for an exploration of the unique contribution of each new set of variables beyond those already accounted for in the previous model. All the models include hierarchical random effects for the study participants to account for repeated measurements. 95% confidence intervals were reported using bootstrapping. Model performances were compared using the Likelihood Ratio Test (LRT) to ensure model parsimony.

All statistical analyses were performed using R software (version 3.6.1) [9]. Linear mixed models were tested using the lme4 package in R [11]. P values for linear mixed models were calculated using the lmerTest package in R [12].

## Results

### Data Summary

In total, 112 users and 27,350 nights were included for TST and MS analyses. The models for the variability of TST and MS used a weekday standard deviation of both measures, which contained 3,682 observations. The average age of the participants was 39.5 years (±9.9 years). 49 were academic staff while 62 were service staff. Figure 1 presents the average values of the four sleep metrics - TST, MS, and their corresponding standard deviations - for each participant included in the analysis.

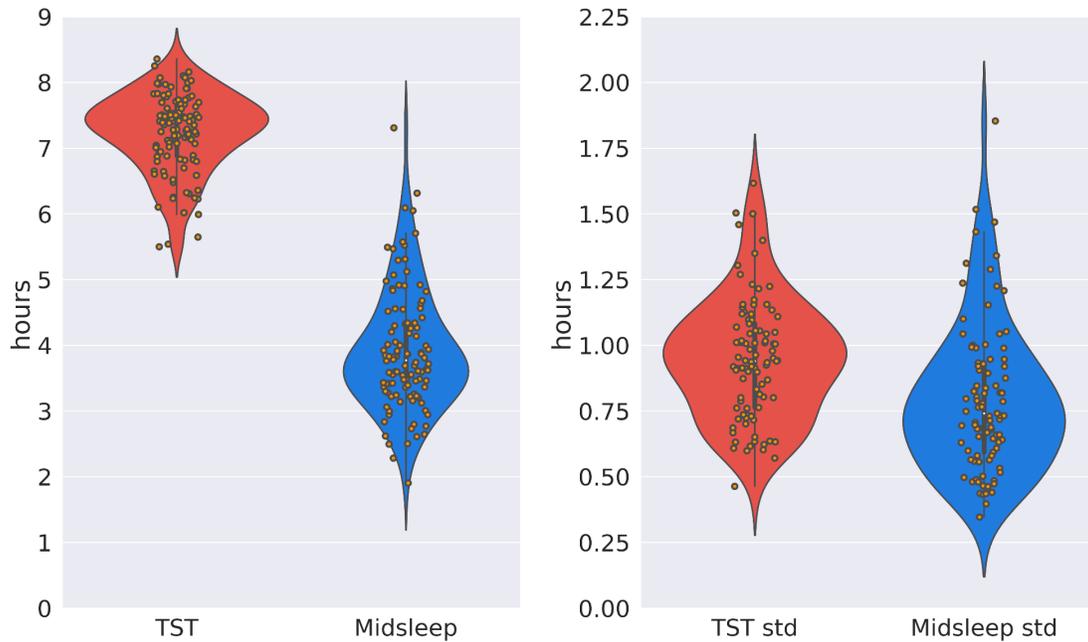

**Figure 1:** TST, MS, and their standard deviations of participants included in the analysis. Each dot represents the participant's mean value for the corresponding metrics.

**Total sleep time**

|  | Model 1 | | | Model 2 | | | Model 3 | | |
|---|---|---|---|---|---|---|---|---|---|
| *Predictors* | Est | CI | p | Est | CI | p | Est | CI | p |
| Age | -0.02 | -0.03 to -0.01 | *.002**  | -0.02 | -0.03 to -0.01 | *.002**  | -0.02 | -0.03 to -0.01 | *.008**  |
| Gender (male) | -0.34 | -0.55 to -0.14 | *<.001*** | -0.34 | -0.55 to -0.14 | *<.001*** | -0.33 | -0.55 to -0.13 | *<.001*** |
| # of children | 0.04 | -0.07 to 0.15 | .51 | 0.04 | -0.07 to 0.15 | .51 | 0.06 | -0.05 to 0.17 | .32 |
| Origin (immigrant) | 0.01 | -0.27 to 0.27 | .97 | 0.01 | -0.27 to 0.27 | .97 | 0.00 | -0.28 to 0.26 | .96 |
| Occupation (service) | 0.36 | 0.11 to 0.60 | *.01* | 0.36 | 0.11 to 0.60 | *.01* | 0.36 | 0.10 to 0.61 | *.006** |
| MEQ | -0.01 | -0.04 to 0.02 | .59 | -0.01 | -0.04 to 0.02 | .60 | 0.00 | -0.04 to 0.03 | .93 |
| Snoozer (Yes) | -0.2 | -0.44 to 0.05 | .11 | -0.2 | -0.44 to 0.05 | .11 | -0.21 | -0.45 to 0.05 | .09 |
| Free day (Yes) | 0.10 | 0.07 to 0.13 | <.001*** | 0.10 | 0.07 to 0.13 | <.001*** | 0.11 | 0.09 to 0.14 | <.001*** |
| Stringency index | | | | 0.005 | 0.003 to 0.007 | <.001*** | 0.004 | 0.002 to 0.006 | <.001*** |
| Day length | | | | -0.01 | -0.01 to -0.01 | <.001*** | -0.009 | -0.012 to -0.006 | <.001*** |
| Steps (x1000) | | | | | | | 0.0 | -0.00 to 0.00 | .39 |
| Midsteps | | | | | | | 0.03 | 0.02 to 0.04 | <.001*** |
| Step entropy | | | | | | | -0.30 | -0.32 to -0.27 | <.001*** |
| **Random Effects** | | | | | | | | | |
| σ2 | 1.13 | | | 1.13 | | | 1.1 | | |
| ICC | 0.19 | | | 0.19 | | | 0.19 | | |
| Marginal $R^2$ / Conditional $R^2$ | 0.055 / 0.234 | | | 0.057 / 0.235 | | | 0.074 / 0.253 | | |
| AIC | 85111.092 | | | 85082.981 | | | 80772.453 | | |

**Table 1:** Estimates of fixed effects from linear mixed effects model predicting TST. Model 1: demographic and occupational variables; Model 2: Model 1+ restriction and seasonal factors; Model 3: Model 2 + physical activity (PA) influences.

We begin by investigating the factors that influence TST, using the three linear mixed models mentioned earlier. Table 1 presents the results of the three models predicting TST. For improved interpretability, the rate of change in TST is measured as the estimate of the predictors multiplied by 60 minutes. In the full model (Model 3), an increase in age by one year was linked to a 1.2 minutes decrease in TST (95% CI -1.8 to -0.6, P=. 008). Considering the gender effect, males were found to sleep 19.8 minutes less than females (95% CI -33.0 to -7.8, *P*<.001). When comparing between occupations, we found that service staff sleep 21.6 minutes more than academic staff (95% CI 6.6 to 36.6, *P*=.006). After adjusting for day length and stringency index, an hour increase in day length was associated with a 0.54 minutes decrease in TST (95% CI -0.72 to -0.36, *P*<.001). However, a one-point increase in the stringency index offset this change by 0.24 minutes (95% CI 0.12 to 0.36, *P*<.001). The full model, which took into account physical activity, showed that a one-unit increase in step entropy corresponded to an 18 minutes decrease in TST (95% CI -19.5 to -16.2, *P*<.001). Moreover, an hour increase in midstep was associated with a 1.8 minute increase in TST (95% CI 1.2 to 2.4, *P*<.001).

The marginal $R^2$ values represented the proportion of variance explained by the fixed effects, and the conditional $R^2$ values illustrated the proportion of variance accounted for by both fixed and random effects. The growth in both $R^2$ values signified that the more intricate models, particularly Model 3, explained more variance in the dependent variable. The Likelihood Ratio Test (LRT) between Model 1 and Model 2 revealed that Model 2 was a significantly superior fit ($\chi^2$=443.7018, *df*=2, *P*<0.001). Furthermore, the LRT between Model 2 and Model 3 demonstrated that Model 3 was a significantly improved fit ($\chi^2$=66.47093, *df*=3, *P*<0.001). The performance of the full model (Model 3) was additionally substantiated by the Akaike Information Criterion (AIC), which was the lowest for Model 3 (AIC=80772.453), demonstrating that it provided the optimal fit for the data.

**Midsleep**

|  | **Model 1** | | | **Model 2** | | | **Model 3** | | |
|---|---|---|---|---|---|---|---|---|---|
| *Predictors* | Est | CI | p | Est | CI | p | Est | CI | p |
| Age | -0.01 | -0.02 to 0.01 | .38 | -0.01 | -0.02 to 0.01 | .46 | 0 | -0.02 to 0.01 | .56 |
| Gender (male) | 0.13 | -0.16 to 0.41 | .39 | 0.12 | -0.17 to 0.40 | .43 | 0.11 | -0.19 to 0.38 | .49 |
| # of children | -0.1* | -0.33 to -0.02 | *.02** | -0.18 | -0.33 to -0.02 | *.02** | -0.16 | -0.31 to -0.00 | *.046** |
| Origin (immigrant) | 0.19 | -0.18 to 0.55 | .32 | 0.18 | -0.20 to 0.53 | .34 | 0.19 | -0.19 to 0.55 | .32 |
| Occupation (service) | -0.17 | -0.51 to 0.17 | .33 | -0.18 | -0.52 to 0.15 | .29 | -0.19 | -0.54 to 0.16 | .27 |
| MEQ | -0.14 | -0.18 to -0.09 | *<.001*** | -0.14 | -0.18 to -0.09 | *<.001*** | -0.14 | -0.18 to -0.09 | *<.001*** |
| Snoozer (Yes) | 0.27 | -0.06 to 0.61 | .09 | 0.29 | -0.04 to 0.63 | .08 | 0.28 | -0.05 to 0.63 | .09 |
| Free day (Yes) | 0.21 | 0.18 to 0.24 | *<.001*** | 0.21 | 0.18 to 0.24 | *<.001*** | 0.22 | 0.18 to 0.25 | *<.001*** |
| Stringency index | | | | 0.02 | 0.02 to 0.03 | *<.001*** | 0.02 | 0.02 to 0.03 | *<.001*** |
| Day length | | | | 0.00 | 0.00 to 0.01 | *.048** | 0.01 | 0.00 to 0.01 | *.002*** |
| Steps (x1000) | | | | | | | -0.01 | -0.02 to -0.00 | *.04** |
| Midsteps | | | | | | | 0.00 | -0.00 to 0.01 | .20 |
| Steps entropy | | | | | | | -0.13 | -0.16 to -0.09 | *<.001*** |
| **Random Effects** | | | | | | | | | |
| σ2 | 1.38 | | | 1.36 | | | 1.36 | | |
| ICC | 0.26 | | | 0.26 | | | 0.27 | | |
| Marginal *R²* / Conditional *R²* | 0.168 / 0.389 | | | 0.178 / 0.400 | | | 0.179 / 0.400 | | |
| AIC | 86990.188 | | | 86573.060 | | | 86540.229 | | |

**Table 2**: Estimates of fixed effects from linear mixed effects models predicting MS. Model 1: demographic and occupational variables; Model 2: Model 1+ restriction and seasonal factors; Model 3: Model 2 + physical activity (PA) influences.

Using the same approach, we provide three linear mixed models to assess the associations between the same set of predictors and MS. The results are presented in Table 2. Using the same approach as above to improve interpretability, the rate of change in MS is measured as the estimate of the predictors multiplied by 60 (minutes). Across the three models, chronotype (MEQ), the number of children, and sleep on a free day consistently emerged as significant factors. In the full model (Model 3), a point increase in the MEQ corresponded with an 8.4 minute decrease in MS (95% CI -10.8 to -5.4, *P*<.001). Sleep on a free day tended to occur 13.2 minutes later (95% CI 10.8 to 15, *P*<.001), compared to a workday. When adjustments were made for the season and restriction policies, we found that MS was delayed by 0.6 minutes (95% CI 0.6 to 1.2,

*P*<.001) following an hour's increase in day length. A point increase in the stringency index also resulted in a 1.2 minutes decrease in MS (95% CI 1.2 to 1.8, *P*<.001). In the full model, which included physical activity variables, a unit increase in steps entropy was associated with a 7.8 minutes earlier MS (95% CI -9.6 to -5.4, *P*<.001). Similarly, an increase in step count was linked with 0.6 minutes earlier MS (95% CI -1.2 to 0.0, *P*=.04).

While the more complex models did not significantly surpass the baseline model in terms of $R^2$, the LRT between Model 1 and Model 2 revealed that Model 2 was a better fit ($\chi^2$=443.7018, df=2, *P*<0.001). Additionally, the LRT between Model 2 and Model 3 demonstrated that Model 3 was a significantly improved fit ($\chi^2$=66.47093, df=3, *P*<0.001). The AIC value for Model 3 was also the lowest (AIC=86540.229), indicating that it provided the best fit for the data.

**Total sleep time variability**

|  | Model 1 | | | Model 2 | | | Model 3 | | |
|---|---|---|---|---|---|---|---|---|---|
| *Predictors* | *Est* | *CI* | *p* | *Est* | *CI* | *p* | *Est* | *CI* | *p* |
| Age | 0.01 | 0.00 to 0.01 | *.01** | 0.01 | 0.00 to 0.01 | *.01** | 0.01 | 0.00 to 0.01 | *.01** |
| Gender (Male) | 0.11 | 0.01 to 0.21 | *.038*** | 0.11 | 0.01 to 0.21 | *.03*** | 0.11 | 0.01 to 0.21 | *.03** |
| # of children | -0.05 | -0.11 to 0.00 | .056 | -0.05 | -0.11 to 0.00 | .052 | -0.06 | -0.11 to -0.00 | *.03** |
| Origin (immigrant) | -0.03 | -0.15 to 0.10 | .56 | -0.03 | -0.15 to 0.09 | .54 | -0.03 | -0.16 to 0.09 | .51 |
| Occupation (service) | -0.17 | -0.28 to -0.05 | *.004*** | -0.17 | -0.28 to -0.05 | *.004*** | -0.16 | -0.27 to -0.04 | *.006*** |
| MEQ | 0 | -0.01 to 0.02 | .55 | 0 | -0.01 to 0.02 | .57 | 0 | -0.01 to 0.02 | .60 |
| Snoozer (Yes) | 0.18 | 0.07 to 0.30 | *.002*** | 0.18 | 0.07 to 0.30 | *.002*** | 0.18 | 0.07 to 0.30 | *.002*** |
| Daylength | | | | 0 | -0.00 to 0.01 | .16 | 0 | -0.00 to 0.01 | .06 |
| Stringency index | | | | 0 | -0.00 to 0.00 | .10 | 0 | -0.00 to 0.00 | .07 |
| Steps | | | | | | | 0 | -0.01 to 0.00 | .08 |

| | | | | | | | | |
|---|---|---|---|---|---|---|---|---|
| Midsteps | | | | | | -0.01 | -0.02 to -0.00 | .02* |
| Steps entropy | | | | | | 0.03 | -0.04 to 0.09 | .37 |
| **Random Effects** | | | | | | | | |
| ICC | 0.14 | | 0.14 | | | 0.14 | | |
| Marginal $R^2$ / Conditional $R^2$ | 0.059 / 0.195 | | 0.060 / 0.194 | | | 0.064 / 0.196 | | |
| AIC | 5458.745 | | 5457.957 | | | 5454.998 | | |

**Table 3:** Estimates of fixed effects from linear mixed effects model predicting MS variability. Model 1: demographic and occupational variables; Model 2: Model 1+ restriction and seasonal factors; Model 3: Model 2 + physical activity (PA) influences.

Table 3 presents the factors predicting the variability in TST. Across the three models, age, gender, the number of children, occupation, and snoozing behavior emerge as significant factors. Every additional year of age corresponded to a 0.01 unit increase in TST variability (95% CI 0.00 to 0.01, $P=.01$). Similarly, male gender was associated with a 0.11 unit increase in TST variability (95% CI 0.01 to 0.21, $P=.038$). Notably, participants with snoozing habits exhibited higher TST variability, increasing by 0.18 units (95% CI 0.07 to 0.30, $P=.002$). Each additional child slightly reduced the variability in TST, a finding that was statistically significance in Model 3 (-0.06, 95% CI -0.11 to -0.00, $P=.032$). Service staff also demonstrated lower TST variability of 0.16 (95% CI -0.28 to -0.05, $P=.006$) in comparison to academic staff. When accounting for physical activity, a decrease of one hour in midsteps correlated with a 0.01 unit increase in TST variability (-0.01, 95% CI -0.02 to -0.00, $P=.028$). The LRT indicated that Model 2 did not provide an improvement over the baseline ($\chi^2=4.78$, df=2, $P=0.09$), however, Model 3 demonstrated a better performance than the baseline model ($\chi^2=13.75$, df=5, $P=0.01$). Despite this improved performance, the improvement in Model 3 was marginal as the R-squared value did not show a significant increase.

### Midsleep variability

|  | Model 1 | | | Model 2 | | | Model 3 | | |
| --- | --- | --- | --- | --- | --- | --- | --- | --- | --- |
| *Predictors* | Est | CI | p | Est | CI | p | Est | CI | p |
| Age | 0.00 | -0.00 to 0.01 | .41 | 0.00 | -0.00 to 0.01 | .41 | 0.00 | -0.00 to 0.01 | .43 |
| Gender (Male) | 0.10 | -0.03 to 0.23 | .12 | 0.10 | -0.03 to 0.23 | .12 | 0.10 | -0.02 to 0.23 | .11 |
| # of children | -0.09 | -0.16 to -0.02 | .01* | -0.09 * | -0.16 to -0.02 | .01* | -0.09 | -0.16 to -0.02 | .01* |
| Origin (immigrant) | -0.05 | -0.19 to 0.11 | .49 | -0.05 | -0.19 to 0.11 | .49 | -0.05 | -0.20 to 0.10 | .45 |
| Occupation (service) | -0.12 | -0.26 to 0.02 | .11 | -0.12 | -0.25 to 0.02 | .11 | -0.12 | -0.25 to 0.03 | .10 |
| MEQ | 0.00 | -0.02 to 0.02 | .96 | 0.00 | -0.02 to 0.02 | .98 | 0.00 | -0.02 to 0.02 | .94 |
| Snoozer (Yes) | 0.20 | 0.06 to 0.35 | .006** | 0.20 | 0.06 to 0.35 | .006** | 0.19 | 0.06 to 0.34 | .008** |
| Daylength |  |  |  | 0.00 | -0.00 to 0.01 | .34 | 0.00 | -0.00 to 0.01 | .21 |
| Stringency index |  |  |  | 0.00 | -0.00 to 0.00 | 0.94 | 0.00 | -0.00 to 0.00 | .81 |
| Steps |  |  |  |  |  |  | 0.00 | -0.01 to 0.01 | .95 |
| Midsteps |  |  |  |  |  |  | -0.02 | -0.04 to -0.00 | .01* |
| Steps entropy |  |  |  |  |  |  | -0.10 | -0.19 to -0.00 | .04* |
| **Random Effects** | | | | | | | | | |
| σ2 | 0.59 | | | 0.59 | | | 0.59 | | |
| ICC | 0.09 | | | 0.09 | | | 0.09 | | |
| Marginal R² / Conditional R² | 0.034 / 0.120 | | | 0.034 / 0.120 | | | 0.037 / 0.121 | | |
| AIC | 8679.371 | | | 8682.369 | | | 8678.443 | | |

**Table 4:** Estimates of fixed effects from linear mixed effects model predicting MS variability. Model 1: demographic and occupational variables; Model 2: Model 1+ restriction and seasonal factors; Model 3: Model 2 + physical activity (PA) influences.

Table 4 presents the factors predicting the variability of MS. Across the three models, the number of children, snoozing behavior, midsteps, and steps entropy emerge as significant factors. For each additional child, the variability of MS was reduced by 0.09 units (95% CI -0.16 to -0.02, $P$=0.05). In all models, being a snoozer correlated with an increase in MS variability. Specifically, snoozers experienced a 0.20 unit increase (95% CI 0.06 to 0.35, $P$=0.01) in MS variability compared to non-snoozers. To better understand the characteristics of snoozers, we ran an additional analysis based on the [Mattingly2022] study. Interestingly, our results revealed that age and chronotype are significant factors in predicting a snoozer. The full results are shown in Appendix 2.

When accounting for physical activity variables in the full model, MS and steps entropy also gained significance. For each hour increase in midsteps, MS variability lessened by 0.02 units (95% CI -0.04 to -0.00, $P$=0.016). Additionally, for each unit rise in steps entropy, MS variability fell by 0.10 units (95% CI -0.19 to -0.00, $P$=0.04). However, the more complex models did not provide a significant improvement over the baseline model, as validated by LRT (Model 2: $\chi^2$=1.00, df=2, $P$=0.60 and Model 3: $\chi^2$=10.92, df=5, $P$=0.06).

## Discussion

### Principal Findings

Research on sleep conducted during the pandemic primarily focuses on the direct disruption effect of lockdowns on sleep. However, little is known about the changes in sleep during the late stage of the pandemic when the restrictive policies were relaxed. Doing so requires a more holistic view of the other factors that have been known to affect sleep apart from the restriction policies, such as individual demographics or seasonal factors. In this study, we used longitudinal data from 112 working adults over one year to explore the relationship between the over-time change in sleep and multiple factors (restriction policies, seasons, physical activity, sociodemographics). Our findings not only reaffirm several known influences on sleep patterns such as gender and age, but they also highlight the significant role of external factors like restriction policies and seasonal changes. It also provides new findings on the implication of a relaxed work schedule and its effects on sleep patterns among different occupations and snoozers.

**Demographic factors**

Past research indicated various epidemiological factors affecting sleep patterns, most notably age, gender, and chronotype. In line with previous studies, we also find an association between age and sleep in which older people tend to sleep less [15, 16]. However, we discover a correlation between older age and higher TST variability, contradicting prior results [17]. The variance in the observed correlations may be attributed to the current study's use of objective sleep measures, whereas [17] relied on self-reported data. On the other hand, no significant association between MS variability and age is found. Regarding gender differences, we show that males exhibited less consistent and overall shorter TST. While the shorter TST among males is well documented [19, 20], the evidence for gender disparity in TST variability is inconsistent. For example, an actigraphy study on a middle-aged cohort showed that females demonstrated greater TST variability than males [21]. In contrast, a survey-based study conducted on university students [22] revealed no difference in TST variability across genders. We also observe that parental duties significantly influence sleep patterns. Parents typically exhibited earlier sleep times, as well as more consistent TST and MS than non-parents. The underlying reasons for these observations remain uncertain, but one hypothesis is that parents' sleep/wake schedule compared to non-parents, as they need to align their sleep patterns with those of their children. While the specific relationship between parenting and sleep pattern variability has not been extensively studied in previous research, the general concept that living with others (cohabitation) can impact sleep patterns by reducing variability in sleep timing and duration has been explored in other studies [23, 63]. This context aims to highlight that factors related to shared living arrangements, such as parenting, can contribute to sleep pattern regularity.

Even though the role of parenting has not been discussed in previous studies, cohabitation has been found to be associated with lower variability in sleep timing and duration [Minors1998].

**Snoozing behavior**

We reveal a higher variability in TST and MS among individuals who identify as 'snoozers'. As working from home becomes the new norm, snoozers have more freedom to adjust their sleep patterns. Interestingly, our findings also suggest that younger individuals and those with an evening chronotype are more likely to identify as 'snoozers', which suggests a potential interplay between age, chronotype, and the habit of snoozing. The inherent sleep-wake patterns of an individual's chronotype might impact their desire to snooze their alarms. Morning types, who naturally wake up earlier, may not find the need to snooze as they align better with societal schedules, compared to evening types.

**Occupational factors**

The nature of one's occupation is a factor that could influence sleep patterns, a classic example being the difference observed between shift workers and non-shift workers [Hulsegge2019, Ganesan2019]. However, less is known about the difference between different roles in academia. We show that academic staff maintain a shorter as well as more variable TST compared to service staff. Moreover, academic staff also exhibit more variable MS compared to service staff. The flexibility, deadline-driven nature of academic schedules may be the main driver behind the irregular sleep pattern. As academics often need to adjust their schedules to meet project deadlines or prepare for lectures, the dynamic nature of their workload can disrupt regular sleep schedules. Moreover, academic work often requires intellectual and creative work which do not conform to a typical 9-5 workday, which further contributes to irregular sleep patterns.

Nonetheless, it is noteworthy that increased variability in sleep patterns might impact overall health and well-being. For example, studies utilizing actigraphy have found that higher TST variability is linked to an increase in depressive symptoms [25, 26]. The implications of these findings become even more relevant in the context of the COVID-19 pandemic. The shift to remote working and learning introduced even greater flexibility for academic personnel. This flexibility could give them more control over their schedules, but it could also mix work and personal life, leading to longer work hours and more irregular sleep schedules.

**Seasonality and restriction policies**

The influence of lockdown measures during the pandemic on sleep patterns is well documented, with increased TST and later MS observed during lockdown periods [1, 4, 27]. In a more detailed analysis using the SI to measure the severity of the lockdowns, [18] showed that a higher SI was correlated with later and more variable MS. However, the effect of season on sleep is often overlooked in past studies. Seasonal factors such as day length have been shown to influence sleep patterns, particularly sleep duration, and timing [28, 13]. Longer daylight hours during the summer might encourage longer waking periods, while shorter days in the winter might lead to extended sleep duration. In southern Finland (the location of our study) where the variance in day length can vary up to 13 hours between summer and winter, these influences could be more noticeable. Our findings confirm that TST decreases when day length increases, while an increase in SI is associated with an increase in TST. Even though the association between an increase in the SI and TST was small (only a 0.24-minute increase in TST per unit increase of SI ), it is noteworthy that the SI typically fluctuated by a degree of 5-10 points, which means that the actual effect could be larger. Similarly, an increase in day length results in a delay in MS, but this delay is further exacerbated by increased SI. Contrary to the findings of [18], our study did not find a correlation between the SI and the variability of MS. However, it's worth noting that there are methodological differences in our approaches. While [18] conducted their correlation measurements on a monthly basis, our analysis was more granular and performed on a weekly level, which could explain the different conclusions. In summary, it is possible that limited exposure to natural daylight induced by reduced mobility could potentially adjust the effect of day length on sleep. Additionally, in our previous research [38], we found a significant correlation between SI and on-site work attendance. Changes in work patterns, especially the shift towards remote work, could further impact sleep timings and duration. For instance, the absence of daily commuting due to remote work could allow individuals to allocate more time to sleep, leading to potential increases in sleep duration.

**Physical activity**

The connection between physical activity (PA) and sleep has been the subject of numerous studies [48, 29, 49]. Although consistent PA is typically recommended for promoting good sleep, it's important to understand that PA is a complex behavior with multiple elements such as duration, timing, and intensity, each potentially influencing sleep [48]. Therefore, we propose that investigations examining the interplay between sleep and PA need to take into account the multifaceted nature of PA.

When considering the timing of PA and its effect on sleep, we find that doing PA later in the day is linked with longer TST and reduced variability in both TST and MS. This aligns with previous research, such as a review by Youngstedt et. al [29], which suggested that engaging in exercise later in the day can be beneficial for sleep. Similarly, a survey study showed that engaging in light- to moderate-intensity workouts early in the evening might impart beneficial effects on sleep [50]. The impact of PA's intensity on sleep could potentially modify the effects of its timing. Sleep hygiene guidelines have suggested that vigorous exercise late in the night might lead to heightened arousal, subsequently impairing sleep quality [51]. However, recent research seems to challenge this convention. For instance, a study conducted under controlled laboratory conditions by Myllymaki et al. [52] found that exercise performed four hours before bedtime did not disturb sleep. Further, a review by Stutz et al. [54] suggested that evening exercise does not adversely impact sleep; however, exercising less than an hour before bedtime could potentially disrupt sleep.

Unlike the aspects of intensity and timing, the impact of how physical activity is spread across the day on sleep has not been investigated in earlier research. Using Shannon entropy to measure the distribution of PA, we find that higher step entropy, indicating a more evenly distributed physical activity throughout the day, is linked to lower TST, but also to earlier MS and reduced TST variability. A potential explanation for this might be that a more evenly distributed physical activity throughout the day could lead to a more consistent arousal and energy expenditure, which could in turn establish a more regular sleep-wake schedule. On the other hand, the reduced TST might be due to the increased wakefulness throughout the day owing to regular PA, which could shorten the duration of sleep.

## Limitations

This work encounters several inevitable limitations. Our study was conducted among the university staff, resulting in a non-representative sample that may introduce bias and a limited sample size. While we tried to control for all the known factors affecting sleep, potentially unaddressed confounding variables can still exist. Additionally, we used consumer-grade wearables for data collection, which, despite their accessibility, may not match the accuracy and reliability of professional-grade equipment. The study was also geographically limited, which restricts the generalizability of our findings to other cultural or social contexts.

## Conclusions

Through a holistic approach, our study provided insights into the changes in sleep patterns and physical activity levels among working adults during the late stage of the COVID-19 pandemic. The flexible working hours during the pandemic have resulted in a corresponding flexibility in sleep patterns among certain occupations and sleep traits, particularly among individuals who self-identified as snoozers. Our findings highlight the significant role that lifestyle habits play in sleep health, especially during unprecedented times like a global pandemic. Moving forward, it is crucial to further investigate the alterations in sleep patterns among diverse populations. Such research will aid in designing workplace policies in the post-pandemic era, taking into account the potential benefits and drawbacks of remote work. One notable advantage to be considered is the increased amount of sleep that workers may experience, which has the potential to enhance overall efficiency and productivity. As we navigate the future of work, understanding the interplay between work arrangements, lifestyle choices, and sleep quality will be essential for promoting optimal well-being and performance in the workforce.

## Acknowledgments

TA acknowledges the support of the MAGICS infrastructure for running the cor:ona study.  We acknowledge the computational resources provided by the Aalto Science-IT project.

## Conflicts of Interest

The authors declare no conflicts of interest.

## Abbreviations

**TST**: Total sleep time
**MS**: Midsleep point
**PA:** Physical activity
**LRT**: Likelihood Ratio Test
**AIC**: Akaike Information Criterion
**PSG**: polysomnography

## Multimedia Appendix 1

| Snoozer | Odds ratio | z | P-value | 95% CI |
|---|---|---|---|---|
| Age | 0.92 | -2.26 | *.02* | 0.86 to 0.99 |
| Gender (male) | 0.45 | -1.13 | .25 | 0.11 to 1.78 |
| Avg. steps (x1000) | 1.06 | 0.57 | .57 | 0.87 to 1.28 |
| Avg. total sleep time | 0.44 | -1.56 | .11 | 0.16 to 1.23 |
| Extraversion | 1.01 | 0.08 | .92 | 0.82 to 1.24 |
| Agreeableness | 1.13 | 0.98 | .32 | 0.89 to 1.44 |
| Conscientiousness | 0.90 | -0.82 | .41 | 0.70 to 1.16 |
| Negative Emotionality | 0.98 | -0.14 | .88 | 0.79 to 1.22 |
| Open Mindedness | 0.94 | -0.42 | .67 | 0.71 to 1.24 |
| Positive Affect | 1.06 | 0.58 | .56 | 0.87 to 1.30 |
| Negative Affect | 1.07 | 0.58 | .56 | 0.86 to 1.32 |
| PHQ-2 | 1.09 | 0.20 | .84 | 0.49 to 2.44 |
| PSQI | 1.00 | -0.002 | .99 | 0.73 to 1.37 |
| MEQ | 0.71 | -3.02 | *.002* | 0.57 to 0.89 |

**Multimedia Appendix 2**

| Staff | Academic | | Service | |
|---|---|---|---|---|
| | TST | MS | TST | MS |
| July | 7.02 (+-1.25) | 4.50 (+-1.63) | 7.52 (+-1.13) | 4.16 (+-1.49) |
| August | 7.13 (+-1.24) | 4.16 (+-1.50) | 7.42 (+-1.11) | 3.72 (+-1.15) |
| September | 7.15 (+-1.15) | 4.06 (+-1.76) | 7.43 (+-1.09) | 3.49 (+-1.47) |
| October | 7.09 (+-1.26) | 4.09 (+-1.80) | 7.45 (+-1.13) | 3.62 (+-1.35) |
| November | 7.13 (+-1.23) | 3.91 (+-1.41) | 7.36 (+-1.06) | 3.51 (+-1.20) |
| December | 7.22 (+-1.33) | 4.28 (+-1.77) | 7.49 (+-1.21) | 3.85 (+-1.41) |

| Staff | Academic | | Service | |
|---|---|---|---|---|
| January | 7.17 (+-1.34) | 4.30 (+-1.62) | 7.53 (+-1.12) | 3.84 (+-1.32) |
| February | 7.14 (+-1.21) | 4.04 (+-1.56) | 7.44 (+-1.10) | 3.66 (+-1.55) |
| March | 7.04 (+-1.27) | 3.87 (+-1.53) | 7.36 (+-1.07) | 3.42 (+-1.33) |
| April | 6.99 (+-1.24) | 3.90 (+-1.30) | 7.38 (+-1.12) | 3.63 (+-1.64) |
| May | 7.01 (+-1.23) | 4.12 (+-1.59) | 7.25 (+-1.10) | 3.63 (+-1.36) |